\documentclass[12pt]{iopart}

\usepackage{iopams}

\newcommand{\<}{\langle}
\renewcommand{\>}{\rangle} 
\newcommand{\txt}{\textstyle}
\newcommand{\dsp}{\displaystyle}
\newcommand{\ri}{{\rm i}}
\newcommand{\rd}{{\rm d}}
\newcommand{\cN}{\ensuremath{\mathcal{N}}}
\newcommand{\bC}{\ensuremath{\mathbb{C}}}
\newcommand{\half} {{\txt \frac{1}{2}}}

\newcommand{\cP}{\ensuremath{\mathcal{P}}}
\newcommand{\cT}{\ensuremath{\mathcal{T}}}

\renewcommand{\Re}{{\rm Re}\,}
\renewcommand{\Im}{{\rm Im}\,}
\newcommand{\nn}{\nonumber \\}

\makeatletter 

\begin{document}
\title[Random matrix ensembles with reduced symmetry]{$2\times 2$ random matrix ensembles with reduced symmetry: From Hermitian to $\cal{PT}$-symmetric matrices}
\author{Jiangbin Gong$^{1,2,3}$ and Qing-hai Wang$^{1}$}
\address{$^{1}$Department of Physics, National University of Singapore, 117542, Singapore}
\address{$^{2}$Centre for Computational Science and Engineering, National University of Singapore, 117542, Singapore}
\address{$^3$ NUS Graduate School for Integrative Sciences and Engineering, 117597, Singapore}
\pacs{05.40.-a, 05.45.Mt, 89.75.-k} 

\begin{abstract}
A possibly fruitful extension of conventional random matrix ensembles is proposed by imposing symmetry constraints on conventional Hermitian matrices or parity-time- ($\cal{PT}$-) symmetric matrices.  To illustrate the main idea, we first study $2\times 2$ complex Hermitian matrix ensembles with $O(2)$ invariant constraints, yielding novel level-spacing statistics such as singular distributions, half-Gaussian distribution, distributions interpolating between GOE (Gaussian Orthogonal Ensemble) distribution and half Gaussian distributions, as well as gapped-GOE distribution. Such a symmetry-reduction strategy is then used to explore $2\times 2$ $\cal{PT}$-symmetric matrix ensembles with real eigenvalues. In particular, $\cal{PT}$-symmetric random matrix ensembles with $U(2)$ invariance can be constructed, with the conventional complex Hermitian random matrix ensemble being a special case. In two examples of $\cal{PT}$-symmetric random matrix ensembles, the level-spacing distributions are found to be the standard GUE (Gaussian Unitary Ensemble) statistics or ``truncated-GUE'' statistics.
\end{abstract}

\maketitle

\section{Introduction}
Random matrix theory (RMT) has found enormous applications in a variety of research areas \cite{mehta,haakebook,lindabook,special}. Predictions of RMT are often widely applicable and it is of fundamental interest to explore its variants and extensions. Two best-known random matrix ensembles are Gaussian Unitary Ensemble (GUE) of complex Hermitian matrices and Gaussian Orthogonal Ensemble (GOE) of real symmetric matrices. Either GUE or GOE can be characterized by $P(H)\rd\mu(H)$, where $\mu(H)$ is the measure of random matrices $H$ and $P(H)$ is the associated probability density. $P(H)\rd\mu(H)$ is assumed to be invariant under certain symmetry groups, called the symmetry invariant group below. The existence of a symmetry invariant group is an important element of RMT, indicating the representation-independence of an ensemble within its symmetry class. That GUE and GOE represents two different universality classes of statistics is exactly because their symmetry invariant groups are different.  As seen below, the concept of symmetry-invariant group becomes even more important as we extend the conventional random matrix ensemble from Hermitian matrices to parity-time- ($\cal{PT}$-) symmetric matrices.

One can also define ``canonical groups'' of transformations for complex Hermitian matrices and for real-symmetric matrices \cite{haakebook}. Adopting the terminology in Ref.~\cite{haakebook}, a canonical group for a complex Hermitian matrix is defined as the largest group of transformations that can maintain the matrix eigenvalues as well as the Hermitian or real-symmetric form. For $N\times N$ complex Hermitian matrices, the canonical group is $U(N)$; whereas for $N\times N$ real-symmetric matrices, the canonical group is $O(N)$. In general, it is often implicitly assumed that the symmetry invariant group of $P(H)\rd\mu(H)$ must be identical with the canonical group.  This assumption does not always reflect a true physical situation and hence can be lifted. For example, if a complex system is subject to a known external influence which breaks the maximal symmetry of the ensemble, then it is more appropriate to model the system by an ensemble with reduced invariant group, e.g., on a subgroup level. Indeed, in earlier studies of the GUE-GOE transition of random complex Hermitian matrices \cite{transition}, each matrix $H$ is decomposed into a real-symmetric matrix plus an anti-symmetric matrix; then based on this decomposition a probability density $P_{\rm{GUE-GOE}}(H)$ is proposed, which is invariant only under an orthogonal group, not a unitary group.

The well-known probability density $P_{\rm{GUE-GOE}}(H)$ for GUE-GOE transitions is just one possible measure to model an ensemble of complex Hermitian random matrices with reduced symmetry. Because $2\times 2$ Hermitian random matrices are just one special case of $2\times 2$ $\cal{PT}$-symmetric random matrices \cite{wcz}, it should be interesting to explore novel random matrix ensembles by proposing non-conventional $P(H)\rd\mu(H)$ with a reduced symmetry invariant group and then extend our approach to $\cal{PT}$-symmetric random matrix ensembles. Our basic procedure is in fact intuitive: Reducing the invariance symmetry of $P(H)\rd\mu(H)$ can be realized by imposing a certain constraint that is invariant under a reduced symmetry. Though our explicit results are based on $2\times 2$ matrices, they can be already relevant to certain physical situations, e.g., they may capture short-ranged statistical properties of large random matrices. Another motivation to work on $2\times 2$ matrices is that the level-spacing distribution functions, which reflects the level repulsion properties of random matrices, can be obtained analytically for $2\times 2$ matrices with elementary tools only. For random Hermitian matrices with reduced invariance symmetry, various interesting results of level-spacing statistics are found, including singular distributions, half-Gaussian distribution, ``gapped-GOE'' distribution, and distributions interpolating between GOE and half-Gaussian distributions.

Using our approach to the treatment of Hermitian random matrix ensembles with reduced invariance symmetry, we move on to construct ensembles of $2\times 2$ $\cal{PT}$-symmetric random matrices. A general statistical measure is deduced from a statistical-independence requirement and then various constraints can be imposed to obtain particular cases with reduced invariance symmetry, yielding interesting level statistics for non-Hermitian random matrix ensembles (such as  a ``truncated-GUE'' distribution). Two important results are as follows. First, the familiar GUE of $2\times 2$ random matrices can be regarded as a symmetry-reduction result from $2\times 2$ $\cal{PT}$-symmetric random matrices. Second, there exists a class of $2\times 2$ non-Hermitian (but $\cal{PT}$-symmetric) random matrix ensembles whose level-spacing distribution is still given by GUE.

Before ending this section, we wish to point out that extending RMT beyond Hermitian matrix ensembles has been of interest for decades \cite{complex1,haakebook}. Thanks to the newly established topic of ${\cal PT}$-symmetric quantum mechanics, this work offers an unforeseen route of extension by exploiting a recent study that gives fully-parameterized $2\times 2$ complex matrices with real eigenvalues \cite{wcz}. It is hoped that more interests in a broader class of random matrix ensembles can be motivated.

\section{General consideration for Hermitian matrices}
A $ 2\times 2$ Hermitian matrix $H$ has four real parameters, 
\begin{equation}
H=\left(
\begin{array}{cc}
a_{11} & a_{12} + \ri b_{12}\\
a_{12} - \ri b_{12} & a_{22} 
\end{array}
\right).
\end{equation}
The natural choice of measure $\rd\mu(H) = \rd a_{11} \rd a_{12} \rd a_{22} \rd b_{12} $ is invariant under the canonical group $U(2)$. Under the same symmetry, there are two obvious invariants 
\begin{equation}
C_1 \equiv \half \Tr H  \quad {\rm and} \quad
C_2 \equiv \half \Tr H^2 - \left(\half \Tr H\right)^2.
\end{equation}
Note that any normalizable and positive function of $C_1$ and $C_2$ can be regarded as a proper distribution function in RMT. To have a specific distribution, one needs to impose constraints other than the symmetry consideration. For example, one may impose the statistical independence of the variables in $H$. In the case of Hermitian random matrices, this leads to the following probability density function
\begin{equation}
P(H) = {\cal N} \exp \left(- A\, \Tr H^2\right),
\label{ph}
\end{equation}
where ${\cal N}$ is a normalization constant. For convenience, we re-parameterize $H$ by using Pauli matrices $\bsigma$ and the identity matrix $\sigma_0$ as
\begin{equation}
H=e \sigma_0 + {\bf X}\cdot\bsigma,
\label{eqn:hermitian}
\end{equation}
where ${\bf X} \equiv (x,y,z)$. Parameter $e$ is a trivial parameter because it just reflects the reference point of the eigenvalues. In this parametrization, the two $U(2)$ invariants are
\begin{equation}
C_1 = e \quad {\rm and} \quad C_2 = x^2+y^2+z^2.
\end{equation}
A natural $U(2)$-invariant measure of $H$ is given by
\begin{equation}
\rd\mu(H) = \rd e\rd x \rd y \rd z.
\label{uh}
\end{equation}
A GUE ensemble can then be defined by use of the $P(H)$ in Eq.~(\ref{ph}) and the $\rd\mu(H)$ in Eq.~(\ref{uh}). For the sake of comparison with other extended cases, we fix $A=\pi/2$ throughout.  Then, after normalizing the total probability to unity, one finds ${\cal N} = \pi/2$ as well as
the following level-spacing distribution
\begin{equation}
P_{\rm GUE}(s)=\frac{\pi}{2} s^2 \e^{- \pi s^2/4},
\label{gue}
\end{equation}
where
\begin{equation}
s \equiv 2\sqrt{C_2} = 2\sqrt{x^2+y^2+z^2}
 \end{equation}
represents the spacing of two eigenvalues. In the three-dimensional parameter space spanned by $(x,y,z)$, $s$ can be interpreted as twice of the distance from the origin.  Because we have fixed $A$, the average level spacing is not unity here.

To reduce the symmetry of $P(H)\rd\mu(H)$, one may reduce the symmetry of either $P(H)$ or $\rd\mu(H)$.
Although in the following we focus on modifications made to $\rd\mu(H)$, they can be equally interpreted as modifications to $P(H)$. Consider then $O(2)$ as a reduced symmetry invariant group.  Under any $O(2)$ transformation, there is an additional invariant, i.e.,
\begin{equation}
C_3 = y.
\end{equation}
Evidently, in general an arbitrary function $f(C_1, C_2, C_3)$ will be invariant under $O(2)$, but not $U(2)$. Hence the following new measure
\begin{equation}
\rd\mu'(H) = f(C_1,C_2,C_3)\rd\mu(H)
\end{equation}
will have a reduced symmetry as compared with $\rd\mu(H)$.

\section{Case studies of random Hermitian matrices}
\subsection{From GUE to GOE}
To illustrate our methodology, we first show how GOE statistics can naturally emerge if we choose $f(C_1,C_2,C_3)=\delta(C_3)=\delta(y)$.  This constraint reduces a Hermitian matrix to a real symmetric form. One then obtains
\begin{eqnarray}
\int P(H)\rd \mu ' (H) &=& {\cal N} \int \e^{-\pi(e^2+x^2+y^2+z^2)} \delta(y) \rd e\rd x \rd y \rd z \nn
&=& {\cal N} \int \e^{-\pi(e^2+x^2+z^2)} \rd e\rd x  \rd z \nn
&=& {\cal N}\frac{\pi}{2} \int_0^\infty s\, \e^{-\pi s^2 /4} \rd s.
\label{goe}
\end{eqnarray}
Note that in obtaining the above result we first carry out an integration over the delta-constraint and then convert one of the remaining integration variables to $s$ (same technique will be used for all the following cases). Using the normalization of total probability, we find ${\cal  N} = 1$. Then Eq.~(\ref{goe}) directly yields the standard GOE distribution of level spacing, i.e.,
\begin{equation}
P_{\rm GOE}(s)=\frac{\pi}{2} s\, \e^{-\pi s^2/4}.
\end{equation}
Interestingly, such a reduction from GUE to GOE is not as trivial as it looks. Because the off-diagonal matrix element  $x-\ri y$ can be also parameterized by $r\e^{-\ri \phi}$, one might na\"ively think that setting the constraint $\delta(\phi)$ will equally reduce GUE to GOE. However, such a $\delta(\phi)$ constraint is inappropriate because $\phi$ is {\it not}\ invariant under the pre-specified symmetry $O(2)$. So even with this new parametrization, one should use $\delta(r\sin\phi)=\delta(y)$ to correctly reduce the symmetry from $U(2)$ to $O(2)$.

\subsection{Planar symmetry}
To demonstrate how our intuitive idea may lead to previously unknown results, we consider here the $O(2)$ invariant function $f(C_1,C_2,C_3)=\delta(C_3-y_0)=\delta(y-y_0)$. In the parameter space, this constraint describes a plane perpendicular to the $y$-axis.  The ensemble statistics is then characterized by
\begin{eqnarray}
\int P(H)\rd \mu'(H)
&=& {\cal N}\int \e^{-\pi(e^2+x^2+y^2+z^2)} \delta(y-y_0) \rd e\rd x \rd y \rd z \nn
&=& {\cal N}\int \e^{-\pi(e^2+x^2+y_0^2+z^2)} \rd e\rd x \rd z \nn
&=& {\cal N}\frac{\pi}{2} \int_{2|y_0|}^\infty s\, \e^{-\pi s^2 /4} \rd s.
\label{ggoe}
\end{eqnarray}
Upon fixing the normalization constant ${\cal N}$, we can directly read out a new level-spacing distribution function from Eq.~(\ref{ggoe}), i.e.,
\begin{equation}
P_{\rm{gGOE}}(s)= \left\{
\begin{array}{ll}
0,& \quad s<2|y_0|;\\ \\
{\dsp \frac{\pi}{2} s \exp\left[- \frac{\pi}{4} \left(s^2-4y_0^2\right)\right] }, &\quad s> 2|y_0|.
\end{array}
\right.
\end{equation}
$P_{\rm{gGOE}}(s)$ can be termed as a ``gapped-GOE'' distribution because it resembles the GOE distribution except that the level spacing must be larger than $2|y_0|$. The gapped feature can be regarded as an extreme situation of level repulsion. Obviously, this result reduces to GOE when $y_0=0$.

Certainly, any distribution of $y$ other than $\delta(y-y_0)$ is still $O(2)$-invariant. So we may superpose many ensembles with different $y_0$ together to form a larger ensemble with the same reduced symmetry. Let us consider the averaging of $y_0$ over the Gaussian distribution
$P(y_0) \rd y_0 =\epsilon\, \e^{-\pi \epsilon^2 y_0^2} \rd y_0$ with $\epsilon>0.$ Then the overall level-spacing statistics $\overline{P}(s)$ is given by a weighted distribution,
\begin{eqnarray}
\overline{P}(s) &\equiv& \left\< P_{\rm{gGOE}}(s)  \right\>_{y_0}\nn 
 &=& \int_{-\infty}^\infty \epsilon\,\e^{-\pi \epsilon^2 y_0^2}   P_{\rm{gGOE}}(s)  \rd y_0\nn 
&=&\frac{\pi}{2} \epsilon  s\, \e^{-\pi s^2 /4}  \int_{-s/2}^{s/2} \e^{\pi (1-\epsilon^2) y_0^2} \rd y_0.
\end{eqnarray}
$\overline{P}(s)$ obtained above is essentially the well-known interpolation between GUE and GOE \cite{transition}.  This becomes obvious if we set $\epsilon=1$, resulting
\begin{equation}
\overline{P}(s) = \frac{\pi}{2} s\, \e^{-\pi s^2 /4} \int_{-s/2}^{s/2}
\rd y_0 = \frac{\pi}{2} s^2 \e^{-\pi s^2 /4},
\end{equation}
which is exactly the $P_{\rm{GUE}}(s)$ obtained in Eq.~(\ref{gue}).  The case of $\epsilon=1$ is special because the ensemble under this precise condition accidentally possesses a $U(2)$ symmetry invariant group. That even the known GOE-GUE distribution can be modeled here further indicates that our approach based on the $O(2)$-invariant function $f(C_1,C_2,C_3)$ is simple but quite general.

\subsection{Cylindrical Symmetry}
We now start to explore various ensembles of $O(2)$ invariance by allowing the function $f(C_1,C_2,C_3)$ to live on different geometrical objects in the $(x,y,z)$ parameter space.  Here we let $f(C_1,C_2,C_3)=\delta(C_2-C_3^2-\rho_0^2)=
\delta(x^2+z^2-\rho_0^2)$, which describes a cylinder invariant to arbitrary rotation about the $y$ axis. As a result, we have
\begin{eqnarray}
\int P(H)\rd \mu'(H)
&=& {\cal N} \int \e^{-\pi (e^2+x^2+y^2+z^2)} \delta(x^2+z^2-\rho_0^2) \rd e\rd x \rd y \rd z \nn
&=& {\cal N}  \int \e^{-\pi (e^2+\rho^2+y^2)} \delta(\rho^2-\rho_0^2)  \rho\rd \rho \rd\theta \rd e \rd y \label{middle} \\
&=& {\cal N} \pi \int_{2 \rho_0 }^\infty \frac{s}{\sqrt{s^2-4\rho_0^2}} \e^{-\pi s^2 /4} \rd s,
\end{eqnarray}
where we introduced the polar coordinates: $x\equiv \rho\cos\theta$ and $z\equiv \rho\sin\theta$.

The normalization condition requires ${\cal N}= \frac{1}{\pi}\e^{\pi \rho_0^2}$. Accordingly
the level-spacing distribution is found to be
\begin{eqnarray}
P (s)= \left\{
\begin{array}{ll}
0,& \quad s< 2\rho_0;\\
{\dsp  \frac{s}{\sqrt{s^2-4\rho_0^2}}  \exp\left[- \frac{\pi}{4} \left(s^2-4\rho_0^2\right)\right] }, &\quad s>2\rho_0.
\end{array}
\right.
\label{cycin}
\end{eqnarray}
This $P(s)$ in Eq.~(\ref{cycin}) has two noteworthy features. First, it is not only gapped but also singular: $P(s)$ diverges
 as $s$ approaches $2\rho_0$.  Second, as $\rho_0\to 0$,  $P(s)\to \e^{-\pi s^2/4}$, which is a half-Gaussian distribution and hence predicts level clustering.  This limit is somewhat expected because
 as $\rho_0\to 0$, both $x$ and $z$ will be fixed at zero and hence $s$ is solely determined by $|y|$, whose distribution is half-Gaussian.


\subsection{Parabolic Symmetry}
Here we consider
\begin{equation}
f(C_1,C_2,C_3)=\delta(x^2+z^2-2\alpha y)+\delta(x^2+z^2+2\alpha y)
\end{equation}
with $\alpha>0$. This hence introduces another invariant measure $\rd\mu'(H)$. In the parameter space, the constraint $f(C_1,C_2,C_3)$ describes two paraboloids with the $y$-axis as their symmetry axis. After some calculations we obtain
\begin{eqnarray}
\int P(H)\rd \mu'(H)
&=& \frac{\cN}{2} \int \e^{-\pi[e^2+(|y|+\alpha)^2 -\alpha^2]} d\theta \rd e \rd y \nn 
&=& \cN \pi \int_0^\infty \frac{s}{\sqrt{s^2 + 4\alpha^2}} \e^{-\pi s^2 /4} \rd s,
\label{p4}
\end{eqnarray}
where we still adopt $x\equiv \rho\cos\theta$ and $z\equiv \rho\sin\theta$ as an alternative parameterization.  The first line of Eq.~(\ref{p4})
indicates that after we confine the ensemble to a parabolic surface, the ensemble can still be interpreted as that generated by three statistically independent variables $e$, $y$, and $\theta$, whose distribution is given by a Gaussian distribution with zero mean, an even distribution patched by two Gaussian tails, and a uniform distribution in $[0,2\pi)$, respectively. In this sense, statistical independence of random variables is again respected.
\begin{equation}
P(s) = \frac{\e^{-\pi\alpha^2}}{ {\rm Erfc} \left(\alpha\sqrt{\pi}\right)}  \frac{s}{\sqrt{s^2 + 4\alpha^2}} \e^{-\pi s^2/4}  .
\label{eq21}
\end{equation}
This distribution represents a smooth interpolation between the GOE distribution $P_{\rm{GOE}}(s)$ and a half Gaussian distribution.
For $\alpha \to \infty$, $P_{\rm{GOE}}(s)$ is recovered, whereas for $\alpha \to 0$, we have $ P(s)\to \e^{-\pi s^2/4}.$  However,
it should be also noted that for any  fixed $\alpha\ne 0$, we always have $P(s\!=\!0)=0$ and hence level repulsion in the regime of small-$s$.  Note also that such an interpolation between level repulsion and level-clustering is much different from that of
the well-known Berry-Robnik distribution~\cite{robnik,prosen}. Based on physical considerations of quantum systems with a mixed classical phase space structure, the Berry-Robnik distribution interpolates between GOE and the Poisson distribution;  whereas our result here is within the framework of random matrix ensembles with certain $O(2)$ symmetry.  In fact, the Poisson distribution never emerges in our calculations.

Another constraint with a parabolic symmetry can be chosen as $f(C_1,C_2,C_3)=\delta(x^2+z^2- \frac{1}{\gamma}y^4)$ with $\gamma>0$.  This constraint describes a surface obtained by rotating a parabola about an axis perpendicular to its symmetry axis. The final level-spacing distribution function is found to be
\begin{equation}
\int P(H)\rd \mu'(H)  = \cN \int_0^{\infty} \frac{s}{\sqrt{s^2+\gamma} \sqrt{\sqrt{s^2+\gamma} -\sqrt{\gamma}}} \e^{-\pi s^2/4} \rd s,
\end{equation}
which is somewhat complicated.  The $P(s)$ distribution here differs from all previous ones in that it always approaches a
nonzero value for $s\rightarrow 0$, thus predicting level-clustering even in the small-$s$ limit.

\subsection{Conic Symmetry}
Here we let $f(C_1,C_2,C_3)=\delta[x^2+z^2-\beta(y+y_0)^2]$ with $\beta>0$, which describes a double cone in the parameter space.  Performing similar
calculations as before, we arrive at
\begin{equation}
\int P(H)\rd \mu'(H) =
\int_{2g|y_0|}^\infty
\frac{s \exp{[-\frac{\pi}{4}(s^2-4g^2 y_0^2)]}}{\sqrt{s^2-4g^2 y_0^2}}  \rd s,
\label{conic}
\end{equation}
where $g\equiv \sqrt{\beta/(1+\beta)}$. This result is very similar to the cylindrical case discussed earlier, so the level-spacing distribution is
also both singular and gapped. When $y_0$ vanishes, we obtain again a half-Gaussian distribution.

\subsection{Discussion}
Our case studies indicate that dramatically different level-repulsion behavior can be obtained for Hermitian random matrix ensembles with an orthogonal symmetry invariant group. The new distributions found here might be useful in some intermediate situations, where some hidden continuous symmetry, albeit smaller than the canonical group, plays a crucial role.

Physical realizations of the ensembles proposed in this section are also possible. Let us first imagine an ensemble of spin-$\frac{1}{2}$ particles embedded in a solid and subject to a magnetic field of strength $B$ along the $y$-direction.  If we further assume that $B$ is a constant and an internal magnetic field is randomly oriented in the $x-z$ plane,
then the $2\times 2$ spin Hamiltonians describing this physical situation form an ensemble with a planar symmetry.  If $B$ is also random, but the internal field strength is fixed, or if there exists an induction mechanism that establishes a linear or quadratic dependence of the internal field strength upon $B$,  then other types of invariance symmetries discussed in this section may be realized.

\section{$\cal{PT}$-symmetric random matrix ensembles}
${\cal PT}$-symmetric matrices are non-Hermitian matrices, but they can generate real eigenvalues due to the underlying invariance under parity and time reversal operations. Recently such systems have attracted a great deal of theoretical and experimental work \cite{pt-review}. Here we aim to study how ${\cal PT}$-symmetric random matrix ensembles might be constructed and explore the associated level statistics. Because a $2\times 2$ $\cal{PT}$-symmetric matrix has two more parameters than a $2\times 2$ Hermitian matrix (detailed below), an interesting conjecture is that a $2\times 2$ Hermitian random matrix ensemble could be just one symmetry-reduced result from a greater symmetry class. If this is true, then more types of random matrix ensembles may be identified.

We start with a generic $2\times 2$ non-Hermitian matrix characterized by eight real parameters,
\begin{equation}
H_{\rm general}=\left(
\begin{array}{cc}
a_{11} + \ri b_{11} & a_{12} + \ri b_{12}\\
a_{21} + \ri b_{21} & a_{22} + \ri b_{22}
\end{array}
\right).
\end{equation}
Consider now similarity transformations, i.e., $GL(2,\bC)$ as the canonical group. Under a similarity transformation, the invariants are
\begin{equation}
C_1 \equiv \half \Tr H \quad {\rm and}\quad
C_2 \equiv \half \Tr H^2 - \left(\half \Tr H\right)^2.
\end{equation}
Note that both $C_1$ and $C_2$ are complex quantities. The real parts and the imaginary parts are independently invariant under $GL(2,\bC)$. The intuitive measure $\rd\mu(H_{\rm general}) = \prod_{i,j=1}^2 \rd a_{ij} \rd b_{ij}$ is also invariant under $GL(2,\bC)$.

To get a specific distribution function for the matrix elements, we need to impose constraints on top of the symmetry consideration. In line with the GUE and GOE cases, here we start from a statistics-independence consideration (which however depends on the parametrization of $H_{\rm general}$). Under our parameterization detailed above, if we request that $P(H_{\rm general})\rd\mu(H_{\rm general})$ describes the parameters in a statistically independent manner, then $P(H_{\rm general})$ is found to be trivial because it depends only on $C_1$ (after some lengthy calculations). For this reason we introduce a new parameterization by expanding $H_{\rm general}$ in Pauli matrices, i.e.,
\begin{equation}
H_{\rm general}=(e + \ri \epsilon)\sigma_0 + ({\bf X} + \ri {\bf R}) \cdot \bsigma,
\end{equation}
where ${\bf X}\equiv (x,y,z)$ and ${\bf R}\equiv (p,q,r)$. With this set of new variables, the above $GL(2,\bC)$ invariants become
\begin{equation}
C_1 = e + \ri \epsilon \quad{\rm and} \quad
C_2 = {\bf X}^2 - {\bf R}^2 + 2\ri {\bf X}\cdot {\bf R},
\end{equation}
and the invariant measure becomes $\rd\mu(H_{\rm general}) =  \rd e \rd \epsilon \rd {\bf X} \rd {\bf R}$. The requirement that this set of new variables should be statistically independent finally leads to
\begin{equation}
P(H_{\rm general}) = \cN P(e) P(\epsilon) \exp\left[ -2A \left({\bf X}^2 - {\bf R}^2 \right) \right],
\end{equation}
where $P(e)$ and $P(\epsilon)$ are arbitrary distribution functions and $A$ is a constant. As a convenient choice, we choose $P(e)$ and $P(\epsilon)$ such that
\begin{eqnarray}
&&P(H_{\rm general})\rd\mu(H_{\rm general}) \nn
&=&  \cN \exp\left[ -A\, \Re\left(\Tr H_{\rm general}^2\right)\right] \rd e \rd \epsilon \rd {\bf X} \rd {\bf R} \nn
&=& \cN \exp\left[ -2A\left(e^2 - \epsilon^2+ {\bf X}^2 - {\bf R}^2\right)\right] \rd e \rd \epsilon \rd {\bf X} \rd {\bf R},
\label{pttrace}
\end{eqnarray}
where ${\cal N}$ is a normalization constant where appropriate. Such type of  $P(H_{\rm general})\rd\mu(H_{\rm general})$ appears naturally from our consideration of statistical independence. At this point 

\noindent $P(H_{\rm general})\rd\mu(H_{\rm general})$ is much different from what is widely used in previous studies of non-Hermitian random matrix ensembles \cite{haakebook,complex1,jain,karr}. In terms of the above parametrization, the previously considered distribution function \cite{haakebook,complex1,jain,karr}
has the form
\begin{eqnarray}
P(H_{\rm non-Hermitian})
 &=& \cN \exp\left[ -A\,\Tr \left( H_{\rm non-Hermitian}^\dag H_{\rm non-Hermitian}\right)\right] \nn
&=& \cN \exp\left[ -2A\left(e^2 + \epsilon^2+ {\bf X}^2 + {\bf R}^2\right)\right]
\label{eqn:otherP}
\end{eqnarray}
Our $P(H_{\rm general})$ is invariant upon $GL(2,\bC)$ transformations and the $P(H_{\rm non-Hermitian})$ in Eq.~(\ref{eqn:otherP}) is only $U(2)$ invariant. To verify this, it is easy to show that the individual length of ${\bf X}$ or ${\bf R}$ is invariant if only $U(2)$ transformations are considered:
\begin{equation}
C_3 = {\bf X}^2 \quad {\rm and}\quad
C_4 = {\bf R}^2.
\end{equation}
Therefore, the $P(H_{\rm non-Hermitian})$ in Eq.~(\ref{eqn:otherP}) can be considered as a result of symmetry reduction by applying the $U(2)$ symmetric constraint  $$f(C_1,C_2,C_3,C_4)=\exp\left[-4A\left(\epsilon^2+{\bf R}^2\right)\right]$$ to Eq.~(\ref{pttrace}).  Certainly, as we will see below, this is only one special example $U(2)$ symmetric constraint.

The probability density $P(H_{\rm general})$ and the invariant measure $\rd\mu(H_{\rm general})$ we proposed above still do not suffice to define an acceptable random matrix ensemble with real eigenvalues. For all eigenvalues to be real, a $2\times 2$ matrix must have the \cP\cT\ symmetry~\cite{wcz}. This is the case under the following conditions:
\begin{equation}
\Im C_1 = \epsilon = 0\quad {\rm and} \quad \Im C_2 = 2{\bf X}\cdot{\bf R}=0.
\end{equation}
Note that these constraints are $GL(2,\bC)$-invariant also, so we are not reducing the invariance symmetry yet. Following exactly the same line as in the Hermitian cases, we implement these two constraints by modifying the invariant measure to the following:
\begin{equation}
\rd\mu(H_{\cP\cT}) = \delta(\epsilon)\delta\left(\frac{{\bf X}\cdot{\bf R}}{\sqrt{\left|{\bf X}^2-{\bf R}^2\right|}}\right)\rd e \rd \epsilon \rd {\bf X} \rd {\bf R}.
\label{ptensemble}
\end{equation}
The denominator in the second $\delta$-function is also $GL(2,\bC)$-invariant and here it is to give a proper dimension.

Equations (\ref{pttrace}) and (\ref{ptensemble}) now define an ensemble of $2\times 2$ ${\cal PT}$-symmetric random matrices. Purely for the sake of further calculations, we now adopt the parametrization similar to the one used in Ref.~\cite{wcz} for ${\cal PT}$-symmetric matrices, namely,
\begin{eqnarray}
H_{\cP\cT}
= e\sigma_0 + \left(\gamma {\bf n}^r + \ri\nu\sin\eta\, {\bf n}^\theta + \ri\nu\cos\eta\, {\bf n}^\varphi\right)\cdot\bsigma,
\end{eqnarray}
with
\begin{eqnarray}
{\bf n}^r &\equiv& (\sin\theta\cos\varphi,\sin\theta\sin\varphi,\cos\theta),\nn
{\bf n}^\theta &\equiv& (\cos\theta\cos\varphi,\cos\theta\sin\varphi,-\sin\theta),\nn
{\bf n}^\varphi &\equiv& (-\sin\varphi,\cos\varphi,0).
\end{eqnarray}
In term of the new variables,
$x\!=\!\gamma\sin\theta\cos\varphi; $
$y\!=\!\gamma\sin\theta\sin\varphi;$
$z\!=\!\gamma\cos\theta;$
$p\!=\!\nu(\sin\eta\cos\theta\cos\varphi - \cos\eta\sin\varphi); $
$q\!=\!\nu(\sin\eta\cos\theta\sin\varphi + \cos\eta\cos\varphi); $
$t\!=\!-\nu\sin\eta\sin\theta, $
with ${\bf X}^2=\gamma^2$ and ${\bf R}^2=\nu^2$. With this parameterization, we have
\begin{eqnarray}
&& \int P(H_{\cP\cT}) \rd\mu(H_{\cP\cT}) \nn
&=& \cN \int \exp\left[ -2A \left(e^2+\gamma^2-\nu^2 \right)\right] \gamma\nu\sin\theta \sqrt{\left|\gamma^2-\nu^2\right|} \, \rd e \rd\gamma \rd\nu \rd\theta \rd\varphi \rd\eta.
\label{eqn:PT}
\end{eqnarray}
To focus on real eigenvalues, we next introduce a constraint such that only ${\cal PT}$-symmetric matrices with real eigenvalues are included, which can be realized by the constraint $\gamma \geq \nu$ \cite{wcz}. We hence introduce a $GL(2,\bC)$-invariant constraint as a step function $\Theta$, i.e.,
\begin{eqnarray}
&& \int P(H_{\cP\cT}) \rd\mu(H_{\cP\cT})\nn
&= & \cN \int \exp\left[ -2A \left(e^2+\gamma^2-\nu^2 \right)\right] \gamma\nu\sin\theta \sqrt{\gamma^2-\nu^2} \Theta\left(\gamma^2 - \nu^2\right) \nonumber \\
 &&\qquad \times\ \rd e \rd\gamma \rd\nu \rd\theta \rd\varphi \rd\eta.
\label{eqn:PTreal}
\end{eqnarray}
With this rather formal expression, a simple analysis indicates that this is still not a normalizable probability distribution ($\cal{N}\rightarrow$ 0). The apparent reason is that the factor 

\noindent$\exp\left[ -2A \left(e^2+\gamma^2-\nu^2 \right)\right]$ in $P(H_{\cP\cT})$ does not decay quickly with increasing $\gamma^2$ so long as $e^2+\gamma^2-\nu^2$ is kept small. A deeper explanation should be linked with the non-compactness of the $GL(2,\bC)$ group, which is closely related to the Lorentz group $O(3,1)$. 
That is, to have a normalizable distribution function, we need to reduce the invariance symmetry to a compact subgroup. In this case, the largest compact subgroup is $U(2)$. This understanding is also confirmed below when we introduce constraints to reduce the invariance symmetry to $U(2)$ (note: this does not mean the matrices themselves have to be complex Hermitian!).

\subsection{From \cP\cT\ symmetry to Hermiticity}

Obtaining Hermitian matrices from \cP\cT-symmetric matrices can be done by a constraint ${\bf R}^2=\nu^2=0$. This constraint is $U(2)$ invariant because it is about the square of the length of ${\bf R}$. Imposing this constraint, we obtain the following $U(2)$-invariant ensemble
\begin{eqnarray}
&& \int P(H_{\cal{PT}})\rd\mu(H_{\cal {PT}})\nn
 &\rightarrow &  \cN \int \exp\left[ -2A \left(e^2+\gamma^2-\nu^2 \right)\right] \gamma\nu\sin\theta \sqrt{\gamma^2-\nu^2} \delta(\nu^2) \rd e \rd\gamma \rd\nu \rd\theta \rd\varphi \rd\eta\nn
&=& \cN' \int \exp\left[ -2A \left(e^2+\gamma^2\right)\right] \gamma^2\sin\theta\, \rd e \rd\gamma \rd\theta \rd\varphi,
\end{eqnarray}
where the $\eta$-parameter becomes irrelevant and can hence be integrated out. Since here the level spacing $s$ is just $2\gamma$, the above expression immediately leads to the standard GUE distribution $P_{\rm{GUE}}(s)$ given in Eq.~(\ref{gue}). Indeed, using our early variables for Hermitian matrices, the expression for  $\int P(H_{\cal PT})\rd\mu(H_{\cal PT})$ is also seen to be identical with $\int P(H_{\rm Hermitian})\rd\mu(H_{\rm Hermitian})$. Thus, a standard Gaussian unitary ensemble of $2\times 2$ Hermitian random matrices is seen to be a special example of our ${\cal PT}$-symmetric random matrix ensemble under a $U(2)$ constraint.

\subsection{General constraint on $\nu$}

Our success in reducing  $P(H_{\cal{PT}})\rd\mu(H_{\cal {PT}})$ to $P(H_{\rm Hermitian})\rd\mu(H_{\rm Hermitian})$ suggests the existence of $U(2)$-invariant ensembles not seen before by imposing more general $U(2)$ constraints. As a very simple example, we may consider a slice of $\nu=\nu_0\ne 0$ of our original ${\cal PT}$-symmetric ensemble to generate an interesting sub-ensemble with again $U(2)$ symmetry. With this new constraint we have
\begin{eqnarray}
&& \int P(H_{\cal{PT}})\rd\mu(H_{\cal {PT}})\nn
& =& \cN \int \exp\left[ -2A \left(e^2+\gamma^2-\nu^2 \right)\right]  \gamma\nu\sin\theta \sqrt{\gamma^2-\nu^2} \Theta (\gamma^2-\nu^2)\delta(\nu-\nu_0) \nonumber \\
 &&\qquad \times\  \rd e \rd\gamma \rd\nu \rd\theta \rd\varphi \rd\eta.
\end{eqnarray}
After carrying out the necessary integrals, one finds that the statistics of the level spacing $s=2\sqrt{\gamma^2-\nu_0^2}$ is again given by $P_{\rm{GUE}}(s)$, i.e., the standard GUE result. This is somewhat unexpected because after all, our ensemble consists of non-Hermitian random matrices!

\subsection{Constraint on $\gamma$}

In exactly the same fashion, we may introduce a constraint about $\gamma$ to generate new ensembles. Because $\gamma$ itself is again $U(2)$-invariant, an arbitrary function of $\gamma$ may be used as a constraint, producing many types of $U(2)$-invariant ensembles in a manner analogous to our previous Hermitian ensembles. Here we consider the simplest case, in which the constraint is given by $\gamma=\gamma_0$.  Then, we can construct the following ensemble,
\begin{eqnarray}
&& \int P(H_{\cal{PT}})\rd\mu(H_{\cal {PT}})\nn
& =&  \cN \int \exp\left[ -2A \left(e^2+\gamma^2-\nu^2 \right)\right] \gamma\nu\sin\theta \sqrt{\gamma^2-\nu^2} \Theta (\gamma^2-\nu^2)\delta(\gamma-\gamma_0) \nonumber \\
 &&\qquad \times\ \rd e \rd\gamma \rd\nu \rd\theta \rd\varphi \rd\eta.
\end{eqnarray}
Integrating out all the free variables under a fixed level spacing $s=2\sqrt{\gamma_0^2-\nu^2}$, we find
\begin{equation}
P_{\rm{tGUE}}(s)= \left\{
\begin{array}{ll}
{\dsp  \cN s^2  \e^{-\pi s^2/4} }, &\quad s<2\gamma_0;\\
0,& \quad s> 2\gamma_0.
\end{array}
\right.
\end{equation}
This level-spacing statistics can be called as a ``truncated-GUE'' distribution because it is exactly the same as the GUE distribution for $s$ less than a threshold value and zero elsewhere.  With this result, it becomes apparent that many interesting level-spacing statistics may emerge if we consider further a distribution of $\gamma$ values. This will not be pursued here.

\section{Conclusion}
In conclusion, by reducing the invariance symmetry of random matrix ensembles via various constraint functions, our straightforward calculations are able to reveal novel features in the level-spacing statistics of $2\times 2$ random matrices, including both Hermitian matrices and ${\cal PT}$-symmetric non-Hermitian matrices. In the Hermitian case, we find singular distributions, half Gaussian distribution, ``gapped-GOE'' distribution, as well as distributions interpolating between GOE and half-Gaussian distributions. Our construction of random $2 \times 2$ Hermitian matrix ensembles is remarkably different from a recent interesting study by Berry and Shukla~\cite{berry}, where intriguing level-spacing distribution functions for generalized Gaussian ensembles of $2\times 2$ real symmetric random matrices are found in the absence of $O(2)$ invariance symmetry. In the non-Hermitian case, our symmetry-reduction approach to ${\cal PT}$-symmetric random matrix ensembles appears to be more general than previous treatments of non-Hermitian random matrix ensembles~\cite{haakebook,complex1,jain,karr} (which can be also regarded as an outcome of symmetry reduction), thereby establishing new types of $2\times 2$ non-Hermitian random matrix ensembles with $U(2)$-invariance and unexpected level-spacing statistics.
 
It is an open question whether results based on $2\times 2$ matrices are relevant to  
${\cal PT}$-symmetric random matrices with a very high dimension. We do not speculate on this challenging issue because at present how to parameterize such matrices is still unclear.  What can be anticipated, however, is that many other types of random matrix ensembles can be considered once the parameterization of high-dimensional ${\cal PT}$-symmetric random matrices are known, with the known GOE and GUE cases as a very small subset of a much broader class of random matrix ensembles.  

\vspace{0.6cm}
\section*{Acknowledgment}
We thank Jiao Wang and Jayendra N.~Bandyopadhyay for interesting comments on the manuscript.  J.G.~is supported by Academic Research Fund Tier I, Ministry of Education, Singapore (grant No. R-144-000-276-112).


\section*{References}
\begin{thebibliography}{90}
\bibitem{mehta} 
M.L.~Mehta, 
{\it Random Matrices}~3rd Ed.~(Elsevier, San Diego, 2004)

\bibitem{haakebook} 
F.~Haake,
{\it Quantum Signatures of Chaos} (Springer-Verlag, New York, 1992)

\bibitem{lindabook} 
L.E.~Reichl,
{\it The Transition to Chaos: Conservative Classical Systems and Quantum Manifestations}, 2nd Ed.~(Springer, New York, 2004)

\bibitem{special} 
P.J.~Forrester, N.C.~Snaith, and J.J.M.~Verbaarschot,
{\it J.~Phys.~A: Math.~Gen.}, {\bf 36}, R1-R10 (2003)

\bibitem{transition} 
A.~Pandey and M.L.~Mehta,
{\it Commu.~Math.~Phys.}  {\bf 87} 449 (1983);
M.L.~Mehta and A.~Pandey,
{\it J.~Phys.~A: Math.~Gen.} {\bf 16} 2655 (1983)

\bibitem{wcz}
Q.-h.~Wang, S.-z.~Chia, and J.-h.~Zhang,
{\it J.~Phys.~A: Math.~Theor.} {\bf 43} 295301 (2010) 

\bibitem{complex1}
J.~Ginibre,
{\it J.~Math.~Phys.} {\bf 6}, 440 (1965)

\bibitem{robnik} 
M.V.~Berry and M.~Robnik,
{\it J.~Phys.~A: Math.~Gen.} {\bf 17} 2413 (1984)

\bibitem{prosen}
T.~Prosen and M.~Robnik,
{\it J.~Phys.~A: Math.~Gen.} {\bf 27} L459 (1994)

\bibitem{pt-review}
C.M.~Bender, 
{\it Rep.~Prog.~Phys.} {\bf 70} {947} (2007) 

\bibitem{jain}
Z.~Ahmed,
{\it Phys.~Lett.~A} {\bf 308} 140 (2003);
Z.~Ahmed and S.~Jain,
{\it Phys.~Rev.~E} {\bf 67} 045106 (2003)

\bibitem{karr}
Y.N.~Joglekar and W.A.~Karr,
{\it Phys.~Rev.~E} {\bf 83} 031122 (2011)

\bibitem{berry}
M.V.~Berry and P.~Shukla,
{\it J.~Phys.~A: Math.~Theor.} {\bf 42} 485102 (2009)

\end{thebibliography}
\end{document}